\documentclass{dhbenelux}

\usepackage{booktabs} 
\usepackage{authblk} 
\usepackage{graphicx} 
\usepackage{hyperref}

\author[1,2]{Myles Joshua Toledo Tan}
\author[2,*]{Panayiotis V. Benos}

\affil[1]{Department of Electrical and Computer Engineering, Herbert
Wertheim College of Engineering, University of Florida, Gainesville, Florida, 32611, United States of America}
\affil[2]{Department of Epidemiology, College of Public Health \& Health Professions and College of Medicine, University of Florida,
Gainesville, Florida, 32610, United States of America}
\affil[*]{Correspondence: \href{mailto:example@domain.com}{pbenos@ufl.edu}}

\title{Addressing Intersectionality, Explainability, and Ethics in AI-Driven Diagnostics: A Rebuttal and Call for Transdiciplinary Action}

\begin{document}

\maketitle
\copyrightstatement

\begin{abstract}
    \noindent {The increasing integration of artificial intelligence (AI) into medical diagnostics necessitates a critical examination of its ethical and practical implications. While the prioritization of diagnostic accuracy, as advocated by \cite{Sabuncu2025-cy}, is essential, this approach risks oversimplifying complex socio-ethical issues, including fairness, privacy, and intersectionality. This rebuttal emphasizes the dangers of reducing multifaceted health disparities to quantifiable metrics and advocates for a more transdisciplinary approach. By incorporating insights from social sciences, ethics, and public health, AI systems can address the compounded effects of intersecting identities and safeguard sensitive data. Additionally, explainability and interpretability must be central to AI design, fostering trust and accountability. This paper calls for a framework that balances accuracy with fairness, privacy, and inclusivity to ensure AI-driven diagnostics serve diverse populations equitably and ethically.}
\end{abstract}

\section{Introduction}

\cite{Sabuncu2025-cy} propose prioritizing diagnostic accuracy over fairness in AI-driven medical diagnostics, asserting that optimizing performance for individual sub-populations should take precedence over achieving equity across groups. While their intent to enhance diagnostic outcomes is commendable, this perspective neglects critical ethical dimensions. Specifically, their approach overlooks the importance of intersectionality, privacy, and the role of transdisciplinary collaboration in shaping equitable and inclusive AI systems.

As engineers and quantitative scientists ourselves, we acknowledge the grave danger in speaking on matters of ethics without incorporating insights from social scientists, ethicists, and practitioners in the fields we aim to transform. It is easy to reduce complex and nuanced issues into quantifiable metrics, stripping them of their human and societal dimensions. In this rebuttal, we call for the authors, and ourselves, to take several steps back, critically reflect on the implications of their propositions, and consider their potential ripple effects across the biomedical engineering community, healthcare systems, and the populations we aim to serve. Moreover, as \cite{Tan2025-ef} highlight, broader determinants of health, such as lifestyle and environmental factors, profoundly shape patient outcomes and healthcare disparities. These elements, alongside demographic and clinical data, must be incorporated into AI diagnostics to create systems that are both precise and reflective of real-world complexities.

This reflection is not merely an intellectual exercise but a moral imperative to ensure our contributions do not inadvertently exacerbate existing inequities. 

\section{The Centrality of Intersectionality in Fairness and Diagnostics}

\cite{Sabuncu2025-cy} treat sub-populations as discrete entities, a reductionist stance that fails to account for the intersecting factors that shape health outcomes. Intersectionality, introduced by \cite{Crenshaw89}, underscores how overlapping identities, such as race, gender, and socioeconomic status, compound experiences of privilege and discrimination. Ignoring these intersections risks producing diagnostic models that reinforce, rather than mitigate, health disparities.

While \cite{Sabuncu2025-cy} mention using sub-populations such as those defined by race, they do not specify how these groups would be delineated. This omission is problematic given that race, as a social construct, lacks a consistent genetic basis and varies significantly across contexts. For example, African populations are more genetically diverse than populations from other continents, yet diagnostic models often treat "race" as a homogenous variable \citep{Tishkoff2004-hz}. Such simplifications risk perpetuating biases and inaccuracies, particularly when genetic and environmental factors interact with social determinants of health.

Moreover, as \cite{Smedley2005-nr} emphasize, race is a social construct with no consistent biological or genetic basis. The traits often associated with racial categories, such as skin color or hair texture, are shaped by a small subset of genes and do not correspond to deeper genetic variation. Using race as a diagnostic proxy perpetuates misconceptions that racial groups represent biologically distinct categories. These misconceptions obscure the structural inequities, such as systemic racism and unequal access to healthcare, that underlie health disparities.

Granularity presents another challenge. As \cite{Sabuncu2025-cy} suggest, defining increasingly fine-grained sub-populations might improve diagnostic accuracy for small groups, but it also exacerbates issues like data scarcity and resource constraints. Moreover, arbitrary categorizations, such as broad racial labels like "Asian" or "Black", fail to capture the diversity within these groups and could lead to inequities in diagnosis and treatment \citep{Yudell2016-rr}.

While \cite{Tan2025-ef} do not emphasize intersectionality, their insights into the integration of lifestyle and environmental factors provide a complementary perspective. Health outcomes are deeply intertwined with determinants such as diet, physical activity, and air quality, which are shaped by systemic inequities and access to resources. For instance, marginalized populations often face environmental risks (e.g., exposure to pollution) and barriers to healthy lifestyles (e.g., access to nutritious food and safe spaces for exercise). Incorporating these determinants into AI systems is essential for addressing the multifaceted nature of health disparities. Diagnostic models that fail to consider these real-world variables risk oversimplifying the factors contributing to disease and misrepresenting the true health challenges faced by diverse populations.

For instance, diagnostic tools optimized for individual sub-populations may fail to recognize how structural inequities compound across intersecting identities. A low-income Black woman, for example, faces unique vulnerabilities due to the convergence of racial, economic, and gendered inequities. Studies like that of \cite{Obermeyer2019-hr} have shown that failing to account for such complexities can result in biased algorithms that systematically underdiagnose or misdiagnose marginalized groups.

Beyond practical performance, the lack of intersectional fairness in AI systems poses ethical and societal risks. Diagnostic tools perceived as inequitable undermine trust in healthcare, especially among historically marginalized communities. This erosion of trust further entrenches healthcare disparities, counteracting the potential benefits of AI-driven advancements. Healthcare systems that ignore these broader determinants risk perpetuating inequities. For example, diagnostic AI optimized without consideration of environmental contexts might overlook disparities in exposure to chronic stressors or pollutants. By situating AI diagnostics within this broader context, as \cite{Tan2025-ef} propose through their \textit{Healthcare 5.0} framework, we can move toward a human-centric, sustainable, and equitable model. By incorporating more nuanced measures, such as intersectional fairness metrics and environmental, lifestyle, and structural determinants of health, diagnostic AI can better address the complexity of real-world disparities. This requires moving away from arbitrary group definitions and embracing multidimensional data integration to reflect patients' lived realities.

\section{Privacy and Security: Beyond Technical Compliance}

\cite{Sabuncu2025-cy} recommend leveraging sensitive attributes like race or gender to optimize diagnostic accuracy. While theoretically promising, this raises profound concerns about privacy, data security, and compliance with legal frameworks like the General Data Protection Regulation (GDPR). These issues are particularly salient in healthcare, where breaches of sensitive data can have devastating consequences.

The use of sensitive attributes increases the risk of re-identification, especially in datasets linked with external sources. For marginalized communities, this risk is amplified by the potential for misuse of their data, leading to discrimination or stigma. Incorporating sensitive attributes without robust safeguards could expose individuals to systemic harms, undermining the very goals of equity and inclusivity.

Moreover, technical compliance with privacy laws is insufficient to address the ethical dimensions of data use. Patients often lack meaningful consent, with limited understanding of how their data will be used or shared. This asymmetry of power between data collectors and individuals underscores the need for transparency, accountability, and stringent governance frameworks to protect patient rights.

The tension between granularity and practicality further complicates this issue. While finer sub-population distinctions may improve the accuracy of AI systems, they also increase the likelihood of privacy breaches and exacerbate inequities in resource allocation. Strategies like differential privacy and federated learning can mitigate these risks, but their implementation must prioritize equity and inclusivity, particularly for underrepresented groups.
\section{Granularity and Practicality}

While \cite{Sabuncu2025-cy} suggest that defining finer-grained sub-populations could improve diagnostic accuracy, this approach presents significant challenges. Increasing granularity exacerbates issues such as data scarcity and computational demands while raising ethical questions about fairness and inclusivity.

\cite{Oh2015-ia} argue that the lack of diversity in biomedical research hinders our ability to address health disparities effectively. For example, most genetic studies have historically focused on populations of European descent, limiting the generalizability of findings to other groups. This lack of representation reinforces biases in diagnostic tools and exacerbates health inequities for underrepresented populations. Expanding diversity in research is critical not only for improving the accuracy of AI systems but also for fostering trust and engagement among marginalized communities. To balance granularity and practicality, AI diagnostics must prioritize diversity in data collection and incorporate variables that reflect the multifaceted nature of health, such as cultural, environmental, and socioeconomic factors.

\section{The Ethical Role of Engineers in AI Development}

As engineers and quantitative scientists, we must critically examine the limits of our expertise in addressing the ethical dimensions of AI. While our technical skills enable us to design and optimize algorithms, we risk reducing complex societal issues to simplistic and quantifiable metrics that fail to capture the full story. This technocratic approach is particularly dangerous in fields like healthcare, where the stakes are profoundly human.

We, as authors of this rebuttal, recognize that we are not immune to these pitfalls. Our expertise does not absolve us of the responsibility to engage with ethicists, social scientists, and healthcare practitioners. Instead, it compels us to seek their insights, ensuring that our work is grounded in the lived realities of the populations we aim to serve. This collaboration is not optional; raher, it is essential for creating AI systems that reflect the values of equity, justice, and human dignity.

Explainability and robust interpretability are critical in this endeavor. AI systems must be designed to provide clear, comprehensible outputs that can be evaluated by non-technical stakeholders. This transparency is vital not only for building trust but also for identifying and mitigating potential biases or unintended consequences. As engineers, we must prioritize the development of interpretable models and actively advocate for their integration into clinical workflows \citep{Doshi-Velez2017-qe}.

\section{Recommendations for an Inclusive and Ethical Framework}

\subsection{Embrace Intersectional Fairness}
AI models must account for intersecting identities to address the compounded effects of structural inequities. This requires leveraging intersectional fairness metrics that balance accuracy and equity across diverse sub-populations \citep{Bowleg2012-od}.

\subsection{Adopt Privacy-First Principles}
Strategies like differential privacy, federated learning, and robust anonymization should be embedded into AI systems to safeguard sensitive data while preserving utility \citep{Vayena2018-tu}.

\subsection{Foster Transdisciplinary Collaboration}
Ethical AI development necessitates collaboration with social scientists, ethicists, and healthcare practitioners to ensure that technical innovations align with societal values and healthcare goals \citep{Floridi2020-vv}.

\subsection{Prioritize Explainability}
Diagnostic tools must be transparent and interpretable, enabling clinicians, patients, and policymakers to understand and evaluate AI-driven decisions. This clarity is essential for building trust and accountability \citep{Doshi-Velez2017-qe}.

\subsection{Promote Participatory AI Development}
Involve marginalized communities in the design and evaluation of AI systems to ensure their needs and perspectives are reflected. This participatory approach fosters equity and inclusivity.

\section{Conclusion}

While \cite{Sabuncu2025-cy} emphasize the importance of diagnostic accuracy, their dismissal of fairness and the broader socio-ethical implications of their framework presents significant risks. heir reliance on poorly defined sub-populations, such as those based on race, oversimplifies the complexity of genetic diversity and social determinants of health. As \cite{Smedley2005-nr} remind us, race is a socially constructed category that lacks biological consistency, and its use in diagnostics risks perpetuating inequities. \cite{Yudell2016-rr} further caution that using race as a genetic proxy reinforces harmful stereotypes, while \cite{Oh2015-ia} highlight the critical need for diversity in biomedical research to ensure equitable health outcomes. 

\begin{figure}[ht]
    \centering
    \includegraphics[width=\linewidth]{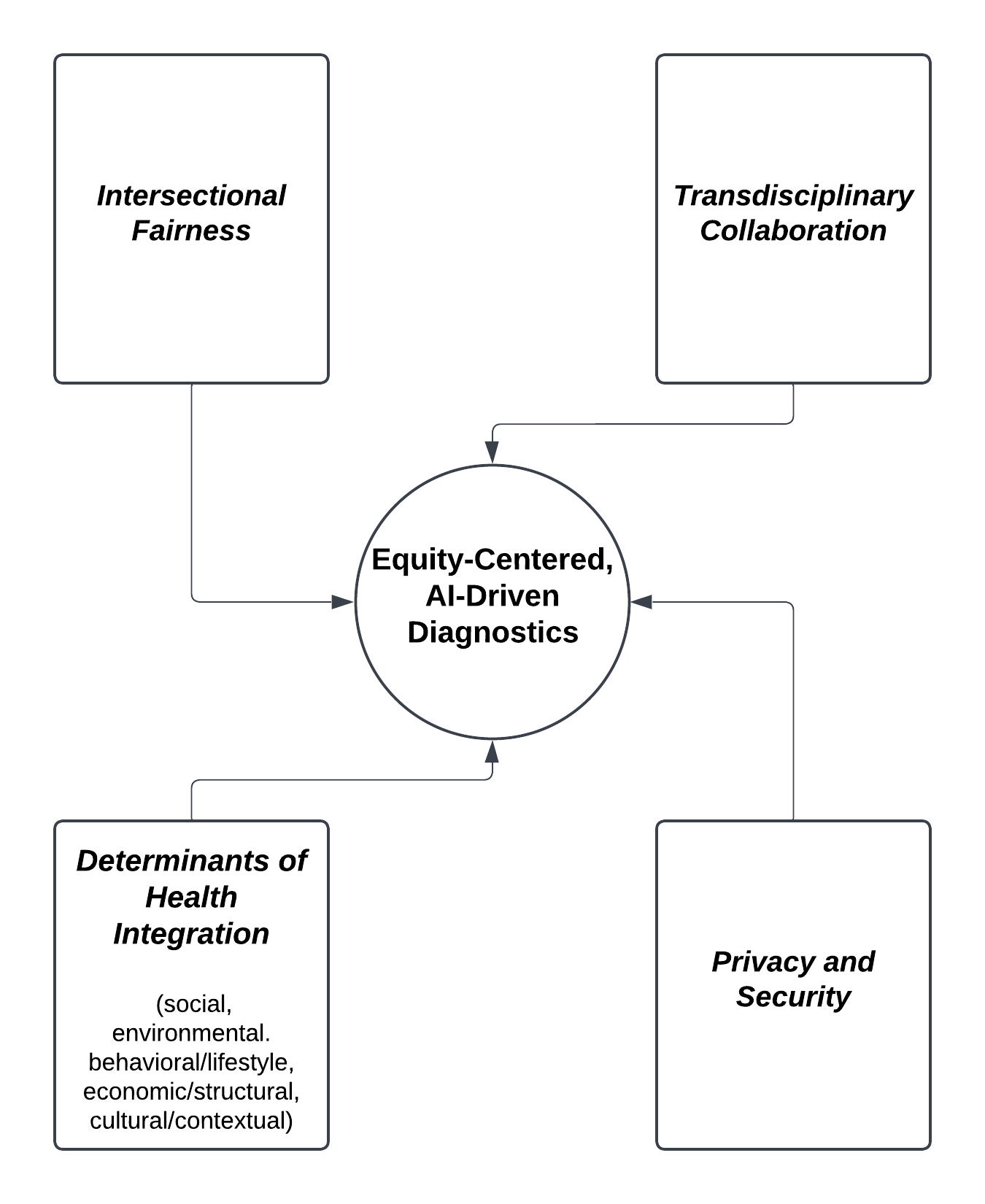}
    \caption{Framework for an Equity-Centered and AI-Driven Diagnostics. This framework illustrates the four pillars essential for achieving equitable and inclusive AI-driven diagnostics. These include Intersectional Fairness, Determinants of Health Integration (social; environmental; behavioral and lifestyle; economic and structural; and cultural and contextual), Privacy and Security, and Transdisciplinary Collaboration. The core goal is to create AI diagnostic systems that prioritize equity while addressing real-world complexities in healthcare.}
    \label{fig:equity-centered-framework}
\end{figure}

As engineers and quantitative scientists, we recognize the dangers of reducing complex ethical issues to numbers and metrics. This acknowledgment calls us, and the authors of the original paper, to critically reflect on our methodologies and their implications. By embracing intersectionality, safeguarding privacy, and fostering transciplinary collaboration, we can create AI systems that not only enhance diagnostic accuracy but also advance equity, trust, and justice in healthcare. The proposed framework, illustrated in Figure~\ref{fig:equity-centered-framework}, summarizes the message of this rebuttal into a cohesive framework diagram. It highlights the four pillars essential to achieving equity-centered AI diagnostics: Intersectional Fairness, Determinants of Health Integration, Privacy and Security, and Transdisciplinary Collaboration. At its core, the framework emphasizes Equity-Centered, AI-Driven Diagnostics as the guiding principle to address real-world complexities and systemic inequities. 

Let us take these steps together, not as isolated experts but as a collective striving toward a shared vision of ethical and inclusive innovation.

\bibliography{references}

\end{document}